\documentstyle[aps,prd,psfig]{revtex}
\begin{document}

\title{Thomas-Fermi-Dirac model  for Low Density\\
Stellar Matter in Presence of a\\
 Strong Quantising  Magnetic Field}

\author{Nandini Nag$^{a)}$\thanks{E-mail:nandini@klyuniv.ernet.in}
and Somenath Chakrabarty$^{a,b)}$\thanks{E-mail:somenath@klyuniv.ernet.in}\\
 a) Department of Physics, University of Kalyani, Kalyani 741 235,
India\\
b) Inter-University Centre for Astronomy and Astrophysics, Post Bag 4\\
Ganeshkhind, Pune 411 007, India}

\date{\today}
\maketitle
\begin{flushleft}
PACS:12.38.Mh, 12.15.Ji, 95.30.Cq, 97.60.Jd
\end{flushleft}

\begin{abstract}
 The effect of strong quantising magnetic field on low density stellar
 matter is investigated  using  Thomas-Fermi-Dirac (TFD) model. The 
 Wigner-Seitz cell  
 structure  is assumed for the low density matter. The significant
 changes in the properties of
 such low density matter  in presence of strong magnetic fields  are 
 discussed. It is seen that the decay time scale for magnetic field
 decreases by at least two orders of magnitude in such model calculation.

\end{abstract}
\vfil\eject
\section{Introduction}
The study of low density stellar matter at the crustal region of a
neutron star or in the case of white dwarf matter in presence of a 
strong magnetic field is
extremely interesting both from the academic as well as from  the
Astrophysical point of view. The recent observational data of magnetars
indicate the possibility of very strong surface magnetic field (up to  
$10^{15}G$)  \cite{dun}
in some pulsars. The observed soft gamma repeaters (SGR) discovered in
BATSE \cite{wood} and  KONUS \cite{maz,hur} experiments and X-ray source
observed by ASCA and RXTE show strong surface magnetic field up to
$10^{15}$G. These objects are called magnetars. They pose a great
challenge to the existing models of magnetic field evolution since they
require a very rapid field evolution in isolated neutron stars. Now the
neutron star crust plays most important role in the evolution of neutron
star magnetic field \cite{kon}.
The neutron star crustal matter (at relatively low density) exhibits
conventional lattice structure, which can be approximated by a regular
arrangement of Wigner-Seitz cells,
 with a positively charged nucleus at the centre surrounded by spherical
distribution of electron cloud. In the case of white dwarfs such
approximation is also valid at low density region. In the non magnetic case,
this kind of
crystalline structure have already been studied in Solid state
physics / Atomic Physics
\cite{qun,mer}. Equation of state for low density stellar matter 
is also obtained
using TFD model in the non magnetic case \cite{bla}. Since the
shell effect of orbital electrons significantly suppresses the statistical
effect, it is not recommended to obtain equation of state for very low
density solids in the laboratory using TFD model (e.g. metallic iron in the
laboratory). On the other hand, in
the case of stellar matter, since the density is high enough, such an
issue does not arise. It was also argued that application of  TFD model
is valid for density  $\leq 10^4$ gm/cc. For higher densities, since electrons 
no longer remain bound within the cells, one uses
Chandrasekhar's ideal gas results for electrons  with  Coulomb lattice 
correction \cite{bla}.

                  If magnetic field of strength $\sim$
  $10^{14}$-$10^{15}$G exists at the outer crust of a neutron star or inside
  a magnetised white dwarf (!) with Wigner-Seitz cell structure of the matter, 
  the
  charge distribution  of the electron cloud within such cells must be affected
  significantly. As a consequence, the size of Wigner-Seitz cells
  will change. This change in cell volume should affect the equation of
  state of low
  density stellar matter and reduces the width of outer crust of neutron
  stars. If such modified equation of state is considered for the
  neutron star crust, it will significantly affect the gross properties including
  the mass-radius relation of the star.
                        To the best of our knowledge, such an
    important issue  has not been discussed before.                      
  
  In this article, we assume that the matter consists of fully ionised
  iron  nuclei and are at rest at the centre
  of Wigner-Seitz cell. Since the matter is at relatively low density,
  the electrons
  surrounding the positive ions are assumed to be non-relativistic. It
  is further assumed that in the microscopic scale the magnetic field is
  constant and is along  z-axis. 
   As we have noticed that the qualitative nature of the
  results do not change if we consider carbon or oxygen instead of iron as the constituent
  of the matter.
  
                  The paper is organised in the following manner. In the
   next  section we  shall develop the basic formalism for
  Thomas-Fermi-Dirac model in presence of a quantising magnetic field.
  We have concluded our results and discussed the future perspective of
  this work at the last section. We have presented a brief outline of
  the derivation 
 of electron-electron exchange interaction (Fock) term in presence of a 
 strong magnetic field in the Appendix A.

   \section{TFD Model in presence of a Strong Magnetic Field}
  
                   In presence of a quantising magnetic field of
  strength  $B$, the electron number density is given by \cite{sc1},
              \begin{equation}
              n_e= \frac{eB}{\pi^2}p_F
                \end{equation}
      where  $ p_F$ is the electron Fermi momentum, and  $e$  is the  magnitude  
      of electronic charge.
      In the Thomas-Fermi model for statistical treatment of atomic  
      structure, it is assumed that within the Wigner-Seitz cell, the electrons
      move in a slowly varying spherically symmetric potential  $V(r)$. Then
      the Fermi energy $\mu$ of an electron is given by,
                  \begin{equation}
                  \mu=-eV(r)+\frac{p_F^2}{2m}
                  \end{equation}
     where $m$ is the electron mass. The Fermi energy $\mu$  is independent of
     $r$, otherwise electron would migrate to a region of smaller $\mu$.
     In  TFD  model    the electron Fermi energy  is given by,
                      \begin{equation}
            \mu=\frac{p_F^2}{2m}-e\phi-u_{ex}(p_F)={\rm{constant}}
                       \end{equation}
where $u_{ex}$ is the exchange part of electron-electron interaction.

     In the non magnetic case \cite{mer,bla},
                    \begin{equation}
                    u_{ex}(p_F)=\frac{e^2}{\pi\hbar}p_F
                    \end{equation}  
    whereas  in the  case of a quantising magnetic field, if all the
    electrons are assumed to be at
    the lowest Landau level, the exchange energy is given by, (see
    Appendix A)
                    \begin{equation}
                    u_{ex}(p_F)=\alpha(1-e^{-\beta p_F})
                    \end{equation}            
    where the parameters  $\alpha$ and $\beta$ are functions of magnetic field
    strength and are given in Table I. Rearranging
    eqn.(3) in  the form (see also \cite{fus}),
                    \begin{equation}
        \frac{p_F^2}{2m}+\alpha e^{-\beta p_F} =\mu^*+e\phi
                    \end{equation}
      where   $\mu^*=\mu+\alpha$ is the modified form of Fermi energy of the
      electron.
      From eqn.(6) one can express Fermi momentum $p_F$ as a function of  
      $\mu^*+e\phi$.
      The numerically fitted functional form is given by a simple power law,
                     \begin{equation}
                     p_F=C(\mu^*+e\phi)^\gamma
                     \end{equation}
     where, $C$ and $\gamma$  are constant parameters for a given magnetic field
     strength.  In Table I, we have shown the variation of $C$ and $\gamma$ with the                                           
     magnetic field strength $B$.
     The potential $\phi$  (which is the  direct interaction term between 
     electron-nucleus and electron-electron) is given by the  Poission's
     equation \cite{bla},
                       \begin{equation}
         \nabla^2\phi=4\pi e n_e+{\rm{nuclear~~ contribution}}
                       \end{equation}
   Since  the nuclear contribution is a delta function about the origin,  
   we can
   omit it for $r>0$ and impose the boundary condition,
                     \begin{equation}
       \lim_{r\rightarrow 0}r\phi(r)=Ze
                     \end{equation}
   Where  $Z$ is the atomic number ($=26$ for iron). The boundary condition 
   at the cell wall of
   radius $r_s$ is that  the electric field vanishes (neutral cell
   condition), which  gives in spherical polar coordinate,
                       \begin{equation}
               \frac{d\phi}{dr}\mid_{r=r_s}=0
                    \end{equation}
   Now,    using the empirically  fitted form of $p_F$ given by
  eqn.(7), and   writing the radial coordinate $r$ in the scaling
  from     $r=ax$,  we have the  Poisson's  equation  (from eqn. (7))
                 \begin{equation}
        \frac{d^2u}{dx^2}=x^{1-\gamma} u^\gamma
                  \end{equation}
  Where 
                \begin{equation}
         \mu^*+e\phi=\frac{Ze^2}{r} u(r)
                \end{equation}
   and                                                 
                \begin{equation}
            a^{3-\gamma}=\frac{\pi\hbar^2 c}{4CB e^{2\gamma+1}Z^{\gamma-1}}
                  \end{equation}
     The boundary condition at the cell boundary  (eqn.(10)) gives,
                    \begin{equation}
                \frac{du}{dx}=\frac{u}{x}
                \end{equation}
          for $x=x_s$.                       
   Since $\gamma <1$ for  the whole range of magnetic field strength 
   ($10^{14}G\leq B\leq 10^{17}$G) of Astrophysical interest, unlike 
   non-magnetic case, the Poisson's equation  (eqn.(11)) does not have   
   singularity at the origin.
   Therefore, the numerical method prescribed by Feynman, Metropolis and
   Teller
 \cite{fey} is not necessary in the quantising magnetic field case. This
qualitative
   change in the form of  Poisson's equation  comes from the  
   modified form of phase space integral of
   electron number density in presence of strong magnetic fields. The 
   non-quantising magnetic field therefore can not
   make any qualitative change in the form of differential eqn.(11) or 
   in other wards in the electron distribution within the cell.
                            Standard  fourth-order Runge-Kutta  method
   has been  used to obtain numerical solution  of
  eqn.(11).  In  this   case the initial value for  the derivative
                           \begin{equation}  
                       \frac{du}{dx}=v_0     
                          \end{equation}
    are chosen by  shooting method to match the boundary condition at the cell  
    surface  for three different magnetic field strengths,
    $B=10^{14}G,10^{15}G$, and $10^{17}G$. The values for $x_s$, the surface 
    scaling parameter are  given in table I for the above three magnetic 
    field strengths.  As we have noticed, 
    the cell radius $r_s=ax_s$  decreases with the increase of magnetic field strength
    and are about an order of magnitude smaller than the non magnetic value 
   \cite{bla}.
    This squeezing of   Wigner-Seitz cell in presence of strong
    quantising
    magnetic field is analogous to the well known magneto-striction
    phenomenon
    observed in classical
    magneto-statics. The variation of $u(x)$ with $x$ for a given magnetic
    field
    strength  is given by the numerically fitted functional
    form
                            \begin{equation}
              u(x)=\frac{u_0}{1+\exp\{\xi(x-x_0)\}}
                           \end{equation}
      where, $u_0$, $\xi$, $x_0$  are constant parameters for a given magnetic
       field strength. The variation of these   parameters with magnetic
       field
       strength are shown in  Table I. In presence of strong quatising magnetic 
       field, the
        variation of $u$ with $x$
       is entirely different from the non magnetic case.  The variation
       is more or less like the radial distribution  of matter in neutron stars.
           Now the equation of state of such cold degenerate low density
 matter is that due to nucleons and electrons  present in the
system.  The pressure contribution mainly comes from the
     electrons. The nuclei are at rest at the centre of each Wigner-Seitz
     cell, therefore we  can ignore their contribution to kinetic pressure 
     of the
     system.  On the other hand, the energy density of the system is mainly 
     dominated by rest mass of the ions. The energy contribution from the 
     electronic sector is
     about five-six orders of magnitude less than the ionic  part,
     therefore one can
     discard the energy contribution in the equation of state from
     electronic part.  
The mass density $\rho$ is simply given by the rest mass of nucleons inside
      the cell,  then we have                                           
                     \begin{equation}
                 \rho=\frac{3Am_B}{4\pi a^3 x_s^3}        
                        \end{equation}
where $m_B=1.66057\times 10^{-24}$g, the effective nucleon mass. In
Table I we have shown the variation of matter density with the magnetic
field strength $B$. Since the radius of Wigner-Seitz cells decrease with
the increase of $B$, there is  an increase in mass density with the increase 
in magnetic field strength. The
expression for kinetic pressure  from electron sector  is given by 
\begin{equation}
P=\frac{eB}{\pi^2} \left [ \frac{p_F^3}{3m} + \alpha \exp(\beta p_F)
\left ( p_F+ \frac{1}{\beta} \right ) - \frac{\alpha}{\beta} \right ]
\end{equation}
where the Fermi momentum  $p_F$ has already been expressed as a function of 
dimensionless surface parameter $x_s$ (see eqn.(7)).
Eqns.(17) and (18) give the equation of state $P=P(\rho)$ in terms of
the surface parameter $x_s$.

In fig.1 we have shown the equation of state of such low density matter
in presence of strong magnetic fields for three different cases:
the upper curve is for
$B=10^{14}G$, middle one is for $B=10^{15}G$, and the lower one is for
$B=10^{17}G$. As we can see from the figure that 
the softness of the matter increases with the increase of magnetic field 
strength. Which also means that the
matter becomes energetically more stable with respect to non-magnetic
case. We have further noticed from the upper and the middle curves that 
at high density
since a large number of Landau levels are populated for electrons, which is
equivalent to the non-quantising picture of external magnetic field,
these two curves coincide at high density. The effect of magnetic field is
completely washed out completely at very high density.
\begin{figure} 
\psfig{figure=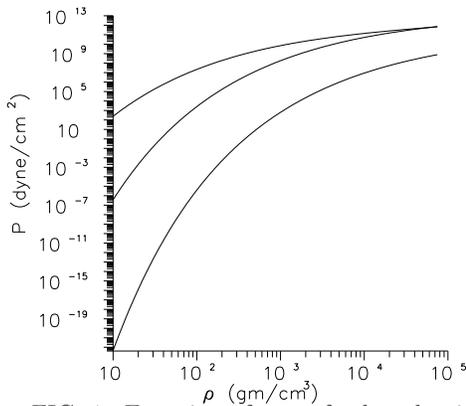,height=0.35\linewidth}
\caption[]{Equation of state for low density neutron star crustal matter
in presence of strong magnetic fields: $B=10^{14}$G (upper curve),
$B=10^{15}$G (middle curve) and $B=10^{17}$G (lower curve)
}
\end{figure}
\section{conclusion}
The outer crust of a neutron star, particularly in the  case of a  strong
magnetic field (magnetars ?) plays a crucial role in the evolution of
pulsar magnetic field.  It is  really a great challenge  to explain
field evolution in these strongly magnetised  objects using existing
models of field evolution.
These objects require a very rapid field evolution. 
Now the TFD model for low density matter in presence of  strong magnetic
fields shows an over all contraction of the outer crust. Since the Ohmic
decay of magnetic field in a conducting material depends on the
thickness of the region, a decrease in width of the outer crust by an
order of magnitude will cause a rapid decay of magnetic field (at least
two orders of magnitude decrease in decay time scale).
The equation of state curves (fig.(1)) indicates that electrons within
the Wigner-Seitz cells are more strongly bound to the positively charged
nuclei in presence of strong quantising magnetic fields than the
non-magnetic (or non-quantising) case.  Such strong binding of electrons
within the cells may decrease the electrical conductivity of the matter.
Which will further reduce the time scale for Ohmic decay of
magnetic field in the outer crust of these strongly magnetised stellar
objects.

\appendix
\section{}
Exchange energy is given by (with $\hbar=c=1$)
\begin{equation}
U_{ex}=\frac{e^2}{2} \sum_{j=1}^Z \int d^3rd^3r^\prime{\frac{1}{\mid \vec r
-\vec r^\prime\mid}} \psi^*(\vec r) \psi(\vec r) \psi(\vec r^\prime)
\psi(\vec r^\prime) 
\end{equation}
For the zeroth Landau level, the wave function $\psi(r)$ for electron is
given by
\begin{equation}
\psi(\vec r)= \frac {1}{\sqrt{L_yL_z}}\left ( \frac{eB}{\pi}\right
)^{1/4} \exp\left [-\frac{eB}{2}(x-\frac{p_y}{eB})^2 \right ] 
\exp[i(p_yy +p_zz)]
\end{equation}
In this particular case the sum over $j$ can be replaced by 
$L_yL_zdp_y^\prime dp_z^\prime$,
where $L_y$ and $L_z$ are the linear dimensions of the box along $y$ and $z$
directions respectively. The volume element $d^3r^\prime= dx^\prime
dy^\prime dz^\prime$.

 Then following Lee \cite{lee},  we have
\begin{eqnarray}
\int dy^\prime dz^\prime \frac{1}{\mid \vec r -\vec r^\prime \mid}
\exp[-i(p_y-p_y^\prime)(y&-&y^\prime) -i(p_z-p_z^\prime)(z-z^\prime)]
 \nonumber \\&=&\frac{4\pi}{2K} \exp\left (-K\mid x-x^\prime \mid \right )
\end{eqnarray}
where $K=\sqrt{(p_y-p_y^\prime)^2+(p_z-p_z^\prime)^2}$.

Similarly $d^3r=dxdydz$. The integral $\int dy dz=L_yL_z$. Then we have
\begin{eqnarray}
U_{ex}&=& \frac{1}{2}\left ( \frac{eB}{\pi} \right ) 4\pi e^2 \int
dp_y^\prime dp_z^\prime dx dx^\prime \frac{1}{2K} \exp \left (-K\mid
x-x^\prime\mid \right )   \nonumber \\ && \exp\left [-\frac{eB}{2} \left \{
\left ( x-\frac{p_y}{eB} \right )^2+
\left ( x-\frac{p_y^\prime}{eB} \right )^2+
\left ( x^\prime-\frac{p_y}{eB} \right )^2+
\left ( x^\prime-\frac{p_y^\prime}{eB} \right )^2
\right \} \right ]
\end{eqnarray}

To evaluate the integrals over $x$ and $x^\prime$, we change the
integration variables to $X$ and $Y$, where $X=x-x^\prime$ and $Y=
(x+x^\prime)/2$.

Now \begin{equation}
\int_{-\infty}^\infty dX \exp\left (-K\mid x\mid \right ) 
\exp\left (-\frac{eB}{2}X^2\right ) =
\sqrt{\frac{2\pi}{eB}} \exp \left (\frac{K^2}{2eB}\right ) 
{\rm{erfc}}\left(
\frac{K}{\sqrt{2eB}}\right )
\end{equation}
where ${\rm{erfc}}(x)$ is the complimentary error function.

Then  we have
\begin{eqnarray}
U_{ex}&=&e^2B \int \frac{1}{K} dp_y^\prime dp_z^\prime dY \sqrt{\frac
{2\pi}{eB}} \exp\left (\frac{K^2}{2eB}\right ) {\rm{erfc}}\left (\frac{K}
{\sqrt{2eB}}\right )  \nonumber \\ && \exp\left [ -\frac{eB}{2} \left ( 4Y^2 +
\frac{2p_y^2}{e^2B^2} +
\frac{2p_y^{\prime 2}}{e^2B^2} -
\frac{4p_yY}{eB}-
\frac{4p_y^\prime Y}{eB} \right )\right ]
\end{eqnarray}

The $Y$ integral is given by
\begin{equation}
\int_{-\infty}^\infty dY \exp \left [ -\frac{eB}{2} \left ( 2Y
-\frac{p_y+p_y^\prime}{eB}\right )^2 \right ]=\sqrt{\frac{\pi}{2eB}}
\end{equation}
Then we have after changing the integration variables from $p_y^\prime$ and
$p_z^\prime$ to $P_y=p_y-p_y^\prime$ and $P_z=p_z-p_z^\prime$
\begin{equation}
U_{ex}=e^2\pi\int dP_y dP_z \frac{1}{\sqrt{P_Y^2+P_z^2}} \exp \left (\frac
{P_z^2}{2eB} \right ) {\rm{erfc}}\left ( \sqrt{\frac{P_y^2+P_z^2} {2eB}}
\right )
\end{equation}
where the limit of $P_y$ is from $-\infty$ to $\infty$ and $P_z$ is from
$0$ to $2p_F$ for $p_z=p_F$.

Again putting $P_y=P_z\tan\theta$, we have 
\begin{equation}
U_{ex}= e^2 \pi \int_0^{2p_F} dP_z \int_0^{\pi/2} \sec\theta ~~d\theta
~~{\rm{erfc}}\left ( \frac{\mid P_z\mid}{\sqrt{2eB}} \sec\theta \right )
\exp \left (\frac{P_z^2}{2eB}\right )
\end{equation}
These double integrals have been evaluated numerically as a function of  Fermi
momentum $p_F$. The fitted functional form of $U_{ex}$ is given by
\begin{equation}
U_{ex}=\alpha [ 1- \exp(-\beta p_F)] \end{equation}
where the parameters $\alpha$ and $\beta$ vary with magnetic field
strength $B$ and are shown in Table I.

\vfil\eject

\noindent {\bf{Table I}}
\bigskip

\begin{tabular} {|l|c|c|r|} \hline 
B (Gauss) & $10^{14}$ & $10^{15}$& $10^{17}$ \\ \hline
$\alpha$ (MeV) &$0.568$& $1.796$& $17.909$\\ \hline
$\beta$ MeV$^{-1}$&$3.412$&$1.067$&$0.109$\\ \hline
$\gamma$ &$0.506$&$0.527$&$0.658$\\ \hline
$C$&$0.973$&$0.870$&$0.386$\\ \hline
$x_s$&$3.096$&$3.170$&$4.404$\\ \hline
$r_s$ (\AA)&$0.402$ &$0.203$& $0.123$ \\ \hline
$v_0$& $-0.938556$& $-0.937365$& $-0.936123$ \\ \hline
$u_0$& $1.633$&$1.651$&$1.944$\\ \hline
$\xi$& $2.097$&$2.071$&$1.755$\\ \hline
$x_0$&$0.213$&$0.204$&$0.031$\\ \hline
$\rho$ (gm/cc)&$72.79$&$572.29$&$962.14$\\ \hline
\end{tabular}

\end{document}